# Using k-alpha emission to determine fast electron spectra using the Hybrid code ZEPHYROS

**Contact** *thomas.white@physics.ox.ac.uk*

**T. G. White, G. Gregori**
*Clarendon Laboratory*
*University of Oxford,*
*Oxford,*
*OX3 0PT*

**A. P. L. Robinson**
*Central Laser Facility*
*Rutherford Appleton Laboratory*
*Harwell Oxford*
*Didcot*
*Oxfordshire, OX11 0QX*

## Introduction

A high intensity laser-solid interaction invariably drives a non-thermal fast electron current through the target. These electrons are accelerated by the radiation field of the laser pulse and can reach energies from a few keV to several MeV. The background resistivity of the solid target or plasma means that the electron current sets up resistive electric fields, a strong return current and magnetic fields. Through a process of collisions and Ohmic heating by the return current the fast electron population loses energy to the surrounding material. On reaching the far side of the target strong sheath fields are set-up which reflect many of the electron back into the target. These processes complicate the dynamics of fast electron transport [1].

Understanding how fast electrons propagate through dense materials is of fundamental interest and has applications relevant to fast ignition schemes and ion acceleration. The return currents also heat solid density material to temperatures from a few to tens of eV creating high energy density states of matter relevant to the study of planetary interiors, warm dense matter and equation-of-state.

A fast electron that has been accelerated through a target produces intense x-ray and VUV emission, primarily through Bremsstrahlung radiation and K-shell ionization in the solid material. The resulting K-shell line emission can be used as an x-ray diagnostic to infer the properties of the fast electron population. Here, we show how the ZEPHYROS hybrid code can be used to infer the spectral temperature, angular divergence and absorbed laser energy of the fast electron distribution from the emitted k-α spectrum. A spectrum of this kind can be obtained experimentally though the use of an absolutely calibrated, imaging, k-alpha spectrometer [2].

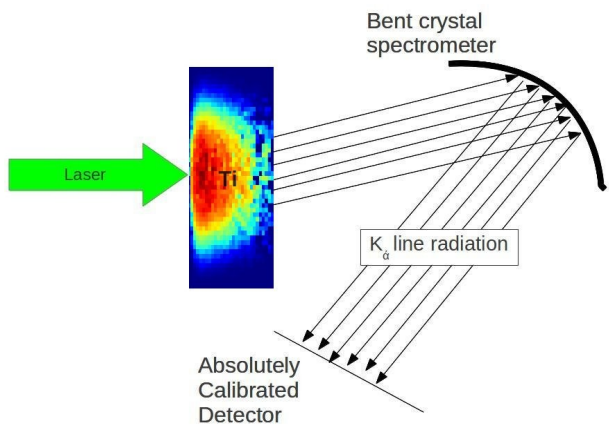

*Figure 1: Possible schematic showing how the k-alpha radiation can be obtained experimentally*

## The code

The ZEPHYROS code is a 3D macroparticle based hybrid code developed by A. P. L. Robinson for the study of electron transport in dense plasmas. The fast electron population is treated as macro particles as in a typical particle-in-cell code while the background electron and ions are treated as a two temperature fluid. The code has many features currently implemented including Bremsstrahlung cooling of the background electrons, electron-ion energy exchange, creation and evolution of magnetic fields and various equation of models for both plasma and solid conditions. Recently, the ZEPHYROS code was upgraded to calculate the k-alpha photon emission rate due to fast-electron-induced k-shell excitation using the algorithm developed by A. G. R. Thomas [3].

The code currently outputs the rate of production of k-alpha photons for each element of the simulation in units of photon number per second per volume. By taking into account radiation transport, the solid angle of a detector and integrating across both the depth of the sample and along a single spatial direction at the back of the target it is possible to obtain the linear intensity of photons on the detector. This number is directly comparable to experimental results obtained from an absolutely calibrated imaging spectrometer.

## Simulation parameters

Here we simulate a $2.5 \cdot 10^{18}$ TW cm$^{-1}$ laser with a 30 micron radius flat-topped spot incident on a 200 micron Ti target for 0.7 ps. The simulation contained 400000 fast electron macro-particles distributed thermally at some characteristic temperature. The simulation was carried out in a 20x30x30 box representing a 200x1800x1800 micron sample. It was run for a time of 6 ps and the k-alpha production was output at 0.2 ps steps. Each simulation was run on the SCARF lexicon cluster operated by the CLF. A single run required a time of approximately 300 seconds included post-processing, and operated on a single core. A total of 18000 simulations were carried out representing a range of fast electron characteristics. This range covered a 3D parameter space of size 30x30x20 representing fast electron spectra temperature (0.02-0.6 eV), laser energy absorption (0.5-15 %) and FWHM divergence angle (63 – 86 degrees).

The figure of merit ($\chi^2$) used for comparison between the simulated spectra and the experimental spectra is achieved using a least mean squares fit between the experimental data and simulation,

$$\chi^2 = \frac{1}{N} \sum_x \left( I_{Simulation}(x) - I_{Experiment}(x) \right)^2$$

Here N is the number of experimental data points, x the position across the target surface and I the linear intensity of k-alpha photons hitting the detector. The minimum of this function represents the conditions where the simulated spectra best matches that observed experimentally.

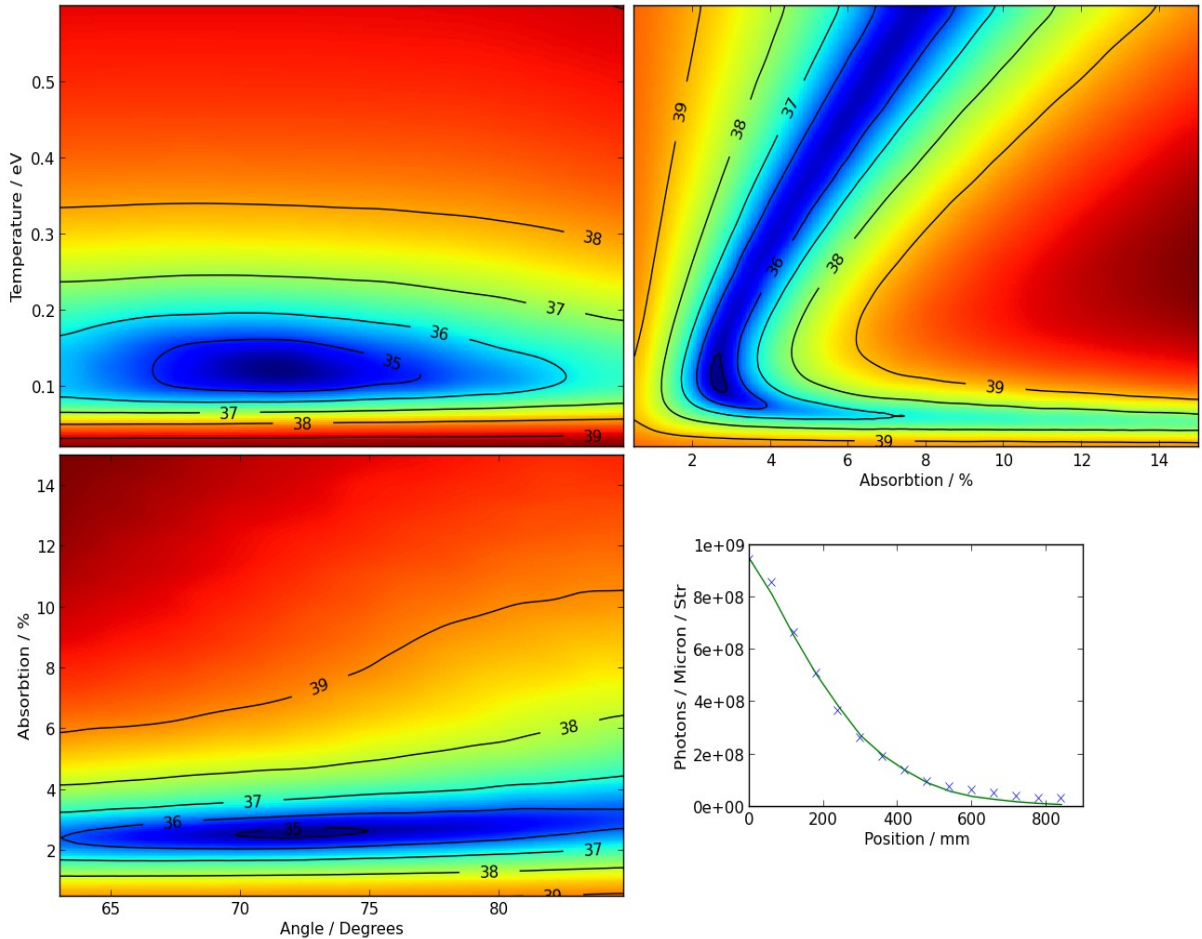

*Figure 2: Log of least mean square fit ($\chi^2$) between simulation and experiment as a function of fast electron properties a) Temperature and Angle b) Temperature and Absorption c) Absorption and Angle. d) Best fitting simulation result (line) and typical experimental k-alpha production (crosses)*

**Results**

To demonstrate this process we have used a typical experimental k-alpha spectrum obtained using the set-up shown in Fig. 1. The lowest figure of merit was obtained when the fast-electron temperature was set to 1.2 eV, the angle to 72 degrees and the absorption fraction of laser light to 2.3%. Plotted in Fig. 2 is the log of this figure of merit shown where one of the electron distribution variables is held constant and the other two are varied. It can clearly be seen that there exists a point in the three dimension space where the figure of merit is minimized. This suggests that these three variables are somewhat decoupled and that this measurement allows all three to be determined to within some degree of accuracy.

Additionally, a region of parameter space where the figure of merit is below a certain value determined by the size of the experimental error bars is traced out. This region contains the predicted properties of the fast electron distribution. Shown in Fig. 2-d is the experimentally obtained k-alpha spectrum (crosses) and the simulation result using the best parameters (line).

**Conclusions**

Utilizing the SCARF lexicon cluster at the CLF facility and the hybrid code ZEPHYROS it is possible to run many thousands of simulations simultaneously to compare with experimental measurements. The addition of k-alpha production to the code along with a small amount of post-processing is one such comparison that could possible give insight into the electron transport and behavior inside a dense target.

Due to the efficient nature of the code it is possible to build up a large parameter space which can then be used to fit experimental data as done here. The agreement between the simulated spectra and experimental results shown in Fig. 2-d both in absolute values and shape is extremely encouraging.

**Acknowledgements**

The author would like to thank A. P. L. Robinson for his support in upgrading the ZEPHYROS code. In addition, the author thanks the SCARF computing services group at the CLF, P. Norreys, and R. Trines for their assistance in gaining access to the SCARF-lexicon computing cluster.